\documentclass{aa}
\usepackage{graphics, epsfig}
\begin{document}

\title{Some astrophysical implications of dark matter and gas profiles in a new galaxy clusters model}

\author{V. F. Cardone\inst{1}
        \and E. Piedipalumbo\inst{2}
        \and C. Tortora\inst{2}}

\offprints{winny@na.infn.it}

\institute{Dipartimento di Fisica ``E.R. Caianiello'', Universit\`a di Salerno, and INFN, Sez. di Napoli, Gruppo Coll. di Salerno, Via S. Allende, 84081 - Baronissi (Salerno), Italy \and Dipartimento di Scienze Fisiche, Universit\`{a} di Napoli, and INFN, Sez. di Napoli, Complesso Universitario di Monte S. Angelo, Via Cinthia, Edificio G - 80126 Napoli, Italy} 

\date{Receveid / Accepted }

\abstract{The structure of the dark matter and the thermodynamical status of the hot gas in galaxy clusters is an interesting and widely discussed topic in modern astrophysics. Recently, Rasia et al. (2004) have proposed a new dynamical model for the mass density profile of clusters of galaxies as a result of a set of high resolution hydrodynamical simulations of structure formation. We investigate the lensing properties of this model evaluating the deflection angle, the lensing potential and the amplification of the images. We reserve particular attention to the structure and position of the critical curves in order to see whether this model is able to produce radial and tangential arcs. To this aim, we also investigate the effect of taking into account the brightest cluster galaxy in the lensing potential and the deviations from spherical symmetry mimicked by an external shear. We also analyze the implication of the gas density and temperature profiles of the Rasia et al. (2004) model on the properties of the X\,-\,ray emission and the comptonization parameter that determines the CMBR temperature decrement due to the Sunyaev\,-\,Zel'dovich effect.

}

\titlerunning{Gravitational lensing and SZ effect from a new galaxy clusters model}

\maketitle

\section{Introduction}

Being the largest bound structures in the universe, galaxy clusters occupy a special position in the hierarchy of cosmic structures in many respects. They can be detected at high redshift because of the presence of hundreds of galaxies and hot X\,-\,ray emitting gas and, therefore, they appear to be ideal tools for studying large scale structure, testing the theories of structure formation and extracting invaluable cosmological information (see, e.g., \cite{BG01,RBN02}).

To first order, galaxy clusters may be described as large dark matter haloes since this component represents up to $80\%$ of the cluster mass. Cluster properties may thus be investigated by means of numerical simulations performed in the standard framework of hierarchical CDM structure formation. The impressive growth in processing speeds of computers in recent years has allowed us to deeply investigate this issue leading to a strong debate about this fundamental topic. While there is a general consensus that relaxed galaxy clusters exhibit a density profile that is well described by a double power law with outer asymptotic slope $-3$, there is still an open controversy about the value of the inner asymptotic slope $\beta$ with proposed values mainly in the range $\sim 1.0 - 1.5$ (\cite{NFW97,TBW97,M98,JS02,P03,N03}). On the other hand, a similar controversy has arisen over the question whether such cusps are indeed observed in galaxies (see, e.g., \cite{S03} and references therein). However, on galaxy scale, the effect of baryonic collapse and astrophysical feedback processes (such as supernova explosions) may alter significantly the dark halo structure thus complicating the interpretation of the observations.

Strongly lensed arcs in galaxy clusters probe the gravitational potential on scales ($r \sim 50 - 100 \ {\rm kpc}$) large enough to avoid baryonic contamination and are thus an ideal tool to investigate this puzzling question. In particular, radial arcs probe the slope of the mass profile at their positions, while tangential arcs constrain the total mass within their radial distance from the cluster centre. Moreover, a measurement of the velocity dispersion essentially fixes the mass divided by the radius even if it is worth stressing that this estimate is reliable only for relaxed clusters. These observables can then be combined to obtain a powerful method to extract information on the cluster structure. 

It is worth noting that the second most important component of a galaxy cluster, namely the gaseous intracluster medium (ICM), is often neglected in numerical simulations. In the usual approach, the ICM distribution is determined {\it a posteriori} from the dark matter density profile imposing the hydrodynamical equilibrium and assuming an isothermal or polytropic equation of state for the gas (\cite{KS01,Asca03}). However, such an approach is somewhat biased since it relies on hypotheses (isothermality and hydrodynamical equilibrium) that do not hold in real galaxy clusters. On the other hand, it is also possible to determine the radial mass profile of both dark matter and ICM directly from simulations explicitly taking into account the gas component. This is the approach followed in a recent paper by Rasia et al. (2004). Using an extended set of high resolution non radiative hydrodynamic simulations and assuming spherical symmetry, these authors have first derived the phase space density of the dark matter particles and then given fitting formulae for the density profile and the velocity dispersion thus allowing them to verify the dynamical equilibrium of the system. Turning then to the hot gas component, they have derived analytic expressions for the density structure, the temperature profile and the velocity dispersion of the ICM without imposing any {\it a priori} hypotheses on the gas dynamical status or its equation of state. In particular, Rasia et al. have shown that the isothermality hypothesis breaks down at distances from the centre larger than $\sim 0.2 R_v$, with $R_v$ the virial radius of the cluster. The Rasia et al. model (hereafter RTM model) presents some peculiarities that make it different from the other models available in literature. Moreover, all the relevant quantities of both the dark matter and gas have been derived in a self\,-\,consistent way free of any bias induced by aprioristic hypotheses on the dynamical state of the system. 

In particular, the knowledge of the gas profile allows one to resort to a completely different (and complementary) observable. With temperature of the order of few keV, the ICM gas is dense and hot enough that clusters are luminous X\,-\,ray sources with the bulk of the X\,-\,rays being produced as bremsstrahlung radiation (\cite{Sar98}). Electrons in the ICM are not only scattered by ions, but may themselves Compton scatter photons of the cosmic microwave background radiation (CMBR) giving rise to the Sunyaev\,-\,Zel'dovich (SZ) effect (see \cite{Bir99} for a comprehensive review). The temperature decrement due to the SZ effect  is able to provide information on the cluster structure, on the motions of galaxy clusters relative to the Hubble flow and on the Hubble flow itself and the cosmological constants that characterize it (\cite{Reese,Mauro03}). 

In order to investigate if the RTM model is a viable one, a direct comparison with the main cluster observables (both from lensing and SZ effect) is needed. As a first step, one has to study the lensing properties of the RTM model and to calculate the SZ effect taking care of the peculiar temperature profile. This is the aim of the present paper.

In Sect. 2 we evaluate the deflection angle and the lensing potential of the spherically symmetric RTM model. Radial and tangential arcs form near the position of the critical curves. Therefore, Sect. 3 is devoted to a detailed investigation of the critical curves structure of the model with a particular emphasis on how these properties depend on the model parameters. The effect of taking into account the contribution of the brightest cluster galaxy to the lensing potential is investigated in Sect. 4, while the impact of deviations from spherical symmetry or tidal perturbations from nearby clusters are mimicked by an external shear and discussed in Sect. 5. Having been proposed recently, the RTM model has to be compared with the previous proposals. In particular, in Sect. 6, we compare some of its lensing properties with those of the NFW model investigating possible systematic errors in the virial mass estimate. The computation of the SZ effect due to the distribution of the ICM in the RTM model is presented in Sect. 7, while the results are compared to the prediction of both the $\beta$ model and the NFW model in Sect. 8 where we also discuss the detectability of RTM clusters in SZ survey. The details of the numerical simulations on which the RTM model is based may induce systematic errors on the main results. Some qualitative comments on this topic are presented in Sect. 9. We summarize and conclude in Sect. 10.

\section{Deflection angle and lensing potential}

Let us adopt a rectangular coordinate system $(x, y, z)$ with origin in the cluster centre and let $(r, \theta, \phi)$ be the usual spherical coordinates. The mass density profile of the RTM model is (\cite{RTM03})\,:

\begin{equation}
\rho(r) = \rho_0 \rho_b \left [ \frac{r}{R_v} \left ( x_p + \frac{r}{R_v} \right )^{1.5} \right ]^{-1}
\label{eq: rhortm}
\end{equation} 
with $\rho_b = \Omega_M \rho_{crit}$ the present day mean matter density of the universe and\,:

\begin{equation}
\rho_0 = \frac{(1 - f_b) \Delta_v}{6 [ (1 + 2 x_p)/(1 + x_p)^{1/2} - 2 x_p^{1/2} ]} 
\end{equation}
where $\Delta_v$ is the virial overdensity specified by the cosmological model and the term $f_b$ is the average baryonic fraction used to properly weight the dark matter component in the cluster. Following Rasia et al. (2004), we set $f_b = 0.097$ as obtained by averaging over their sample of simulated clusters.

The RTM model is fully characterized by two parameters, namely the dimensionless scale radius $x_p$ (or the concentration $c_{RTM} = 1/x_p$) and the virial radius $R_v$. However, it is more convenient to express $R_v$ in terms of the virial mass (i.e. the total cluster mass) $M_v$ using the following relation\,:

\begin{equation}
R_v = \left ( \frac{3 M_v}{4 \pi \Delta_v \rho_b} \right )^{1/3} \ .
\label{eq: rvmv}
\end{equation}
We will assume that the model is spherically symmetric so that all the lensing quantities will depend only on the projected radius $R = (x^2 + y^2)^{1/2}$. This is the same approximation used in Rasia et al. (2004) to obtain the density profile in Eq.(\ref{eq: rhortm}). While useful in the computations, this approximation is not a serious limitation to our analysis since the results for the circular case may be immediately generalized to flattened models by means of numerical integration (\cite{Schramm,K01}). Moreover, we will investigate later the impact of deviations from circular symmetry by adding a shear term to the lensing potential. However, we stress that the results for the circularly symmetric models allow us to obtain a picture of the main properties of the RTM model as a lens.

As a first step to investigate the lensing properties of the RTM model, we have to evaluate the corresponding surface mass density. Starting from the definition\,:

\begin{displaymath}
\Sigma(x, y) = \int_{-\infty}^{\infty}{\rho(x, y, z) dz}
\end{displaymath}
and using a convenient transformation to spherical coordinates, we get\,:

\begin{eqnarray}
\Sigma(\xi) & = & 2 \ \rho_0 \ \rho_b \ R_v \int_{0}^{\pi/2}{\frac{1}{\sin{\theta}} \left ( x_p + \frac{\xi}{\sin{\theta}} 
\right )^{-1.5} d\theta} \nonumber \\ 
~ & = &  \Sigma_v \ \times \ \frac{{\cal{S}}(\xi, x_p)}{{\cal{S}}(1, x_p)} 
\label{eq: sigmartm}
\end{eqnarray}
with $\xi = R/R_v$ and $\Sigma_v$ the surface density at the virial radius given by\,:

\begin{equation}
\Sigma_v \equiv \Sigma(\xi = 1) = \sqrt{\frac{8}{\pi}} \ \rho_0 \ \rho_b R_v {\cal{S}}(1, x_p) 
\label{eq: defsigmav}
\end{equation}
and we have defined the function\,:

\begin{eqnarray}
{\cal{S}} & = & \xi^{-5/2} 
\left \{ \left [ \Gamma\left( \frac{3}{4} \right ) \right ]^2 
{_2F_1 \left [ \left \{ \frac{3}{4}, \frac{3}{4} \right \}; \left \{ \frac{1}{2} \right \}; \frac{x_p^2}{\xi^2}, \right ]} \xi  - \right . \nonumber \\
~ & ~ & \left . - 2 x_p \left [ \Gamma\left( \frac{5}{4} \right ) \right ]^2 
{_2F_1 \left [ \left \{ \frac{5}{4}, \frac{5}{4} \right \}; \left \{ \frac{3}{2} \right \}; \frac{x_p^2}{\xi^2}, \right ]} \right \} \ . 
\label{eq: defesse}
\end{eqnarray}
In the previous equation, $\Gamma(\zeta)$ is the actual $\Gamma$ function and ${_pF_q[\{a_1, \ldots, a_p\}, \{b_1, \ldots, b_q\}, \zeta)}$ is the generalized hypergeometric function\footnote{We use the same notation for the generalized hypergeometric function as in the {\it Mathematica} package.} (\cite{GR80}). 

\begin{figure}
\centering
\resizebox{8.5cm}{!}{\includegraphics{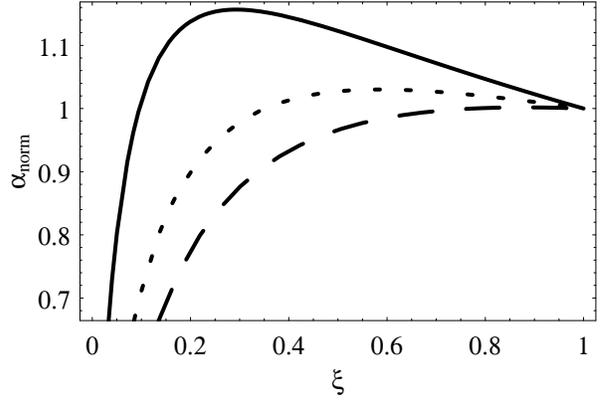}}
\hfill
\caption{The scaled deflection angle $\alpha_{norm} \equiv \alpha/\alpha_v$ vs the dimensionless radius $\xi$ for three values of the dimensionless scale radius $x_p$ of the RTM model, i.e. $x_p = 0.1$ (solid), $x_p = 0.2$ (short dashed), $x_p = 0.3$ (long dashed).} 
\label{fig: alphavsx}
\end{figure}

Having obtained the surface mass density, it is now straightforward to compute the deflection angle. Because of the circular symmetry in the lens plane, the deflection angle is purely radial and its amplitude is (\cite{SEF})\,:

\begin{equation}
\alpha = \frac{2}{R} \int_{0}^{R}{\frac{\Sigma(R')}{\Sigma_{crit}} R' dR'}
\label{eq: defalpha}
\end{equation}
with $\Sigma_{{\rm crit}} = c^2 D_s/4 \pi G D_l D_{ls}$ the critical density for lensing and $D_s$, $D_l$, $D_{ls}$ are the angular diameter distances between observer and source, observer and lens and lens and source, respectively. Inserting Eq.(\ref{eq: sigmartm}) into Eq.(\ref{eq: defalpha}), we get\,:

\begin{equation}
\alpha(\xi) = \alpha_v \times \frac{{\cal{F}}(\xi, x_p)}{{\cal{F}}(1, x_p)}
\label{eq: alphartm}
\end{equation}
with $\alpha_v$ the deflection angle at the virial radius given by\,:

\begin{equation}
\alpha_v \equiv \alpha(\xi = 1) = \frac{4 \Sigma_v R_v}{\Sigma_{crit}} \frac{{\cal{F}}(1, x_p)}{{\cal{S}}(1, x_p)} 
\label{eq: alphav}
\end{equation}
and we have introduced the function\,:

\begin{eqnarray}
{\cal{F}} & = & \frac{1}{\xi} \left \{ \left [ \Gamma\left(\frac{3}{4} \right ) \right ]^2 
{_2F_1 \left [ \left \{ - \frac{1}{4}, \frac{3}{4} \right \}; \left \{ \frac{1}{2} \right \}; \frac{x_p^2}{\xi^2}, \right ]} \xi^{1/2}  + \right . \nonumber \\
~ & ~ & + 2 x_p \left [ \Gamma\left(\frac{5}{4} \right ) \right ]^2 
{_2F_1 \left [ \left \{ \frac{1}{4}, \frac{5}{4} \right \}; \left \{ \frac{3}{2} \right \}; \frac{x_p^2}{\xi^2}, \right ]} \xi^{-1/2}  - \nonumber \\
~ & ~ & \left . - \frac{}{} \sqrt{2 \pi x_p} \ \right \} \ .
\label{eq: defeffe}
\end{eqnarray}
It is interesting to observe that only the parameter $x_p$ of the RTM model determines the behaviour with the dimensionless radius $\xi$ of the scaled deflection angle $\alpha/\alpha_v$, while the virial radius $R_v$ (and thus the total mass $M_v$) enters only as a scaling factor. As an example, Fig.\,\ref{fig: alphavsx} shows $\alpha/\alpha_v$ for three values of $x_p$. This plot may be qualitatively explained as follows. Lower values of $x_p$ correspond to higher values of the concentration $c_{RTM}$ and hence to more mass within a fixed radius. Since, for a given position in the lens plane, the deflection angle scales with the projected mass within $\xi$, it turns out that $\alpha/\alpha_v$ is higher for lower values of $x_p$ as Fig.\,\ref{fig: alphavsx} shows. 

\begin{figure}
\centering
\resizebox{8.5cm}{!}{\includegraphics{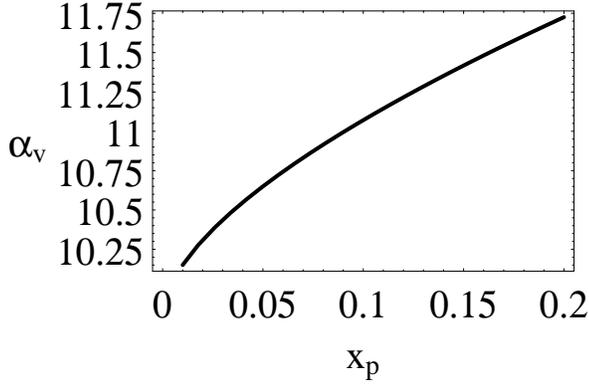}}
\hfill
\caption{The deflection angle at the virial radius $\alpha_v$ vs the dimensionless scale radius $x_p$ of the RTM model. See the text for the values of the other parameters.}
\label{fig: alphavvsxp}
\end{figure}

This qualitative discussion also explains the behaviour of $\alpha_v$ with $x_p$ that is shown in Fig.\,\ref{fig: alphavvsxp}. A comment is in order here to understand how this plot has been obtained. Eq.(\ref{eq: alphav}) shows that $\alpha_v$ depends on the cluster parameters $(M_v, x_p)$, the lens and source redshift $(z_l, z_s)$ and the background cosmological model. We adopt a flat $\Lambda$CDM model with $(\Omega_M, \Omega_{\Lambda}, h) = (0.3, 0.7, 0.72)$ giving $\Delta_v = 324$ (\cite{ECF96}). We set $(z_l, z_s) = (0.313, 1.502)$ as for the real cluster lens MS2137-23 (Sand et al. 2002, 2004). Unless otherwise stated, the same values for the cosmological parameters and the lens and source redshift will be used throughout the paper. To obtain the plot in Fig.\,\ref{fig: alphavvsxp}, we have fixed $M_v = 7.5 \times 10^{14} \ {\rm M_{\odot}}$, but the results for other values of $M_v$ may be easily scaled observing that $\alpha_v \propto \Sigma_v R_v \propto R_v^2 \propto M_{v}^{2/3}$. 

Let us now derive the lensing potential $\psi(R)$ for the RTM model. To this aim, one should solve the two dimensional Poisson equation (\cite{SEF})\,:

\begin{equation}
\nabla^2 \psi = 2 \kappa
\label{eq: poisson}
\end{equation}
with $\kappa = \Sigma/\Sigma_{crit}$ the convergence. However, because of the circular symmetry in the lens plane, it is also\,:

\begin{equation}
\alpha(R) = \frac{d\psi}{dR} \rightarrow \psi(R) = \int{\alpha(R) dR} \ .
\label{eq: alphapsi}
\end{equation}
Inserting Eq.(\ref{eq: alphartm}) into Eq.(\ref{eq: alphapsi}) and integrating, we find\,:

\begin{equation}
\psi(\xi) = \psi_v \times \frac{{\cal{P}}(\xi, x_p)}{{\cal{P}}(1, x_p)}
\label{eq: psirtm}
\end{equation}
with\,:

\begin{equation}
\psi_v \equiv \psi(\xi = 1) = \alpha_v R_v {\cal{P}}(1, x_p) \ , 
\label{eq: psiv}
\end{equation}

\begin{eqnarray}
{\cal{P}} & = & - \frac{\Gamma\left ( - \frac{1}{4} \right ) \Gamma\left ( \frac{3}{4} \right )}{2} 
{_2F_1 \left [ \left \{ - \frac{1}{4}, - \frac{1}{4} \right \}; \left \{ \frac{1}{2} \right \}; \frac{x_p^2}{\xi^2}, \right ]} \xi^{1/2}  - \nonumber \\
~ & ~ & - x_p  \Gamma\left ( \frac{1}{4} \right ) \Gamma\left ( \frac{5}{4} \right ) 
{_2F_1 \left [ \left \{ \frac{1}{4}, \frac{1}{4} \right \}; \left \{ \frac{3}{2} \right \}; \frac{x_p^2}{\xi^2}, \right ]} \xi^{-1/2}  - \nonumber \\
~ & ~ & - 2 \sqrt{2 \pi x_p} \ln{\xi} \ .
\label{eq: defpi}
\end{eqnarray}
Finally, let us consider the lens equations. Adopting polar coordinates $(R, \vartheta)$ in the lens plane with $\vartheta$ measured counterclockwise from North, the time delay of a light ray deflected by the gravitational field of the cluster lens is\,:

\begin{eqnarray}
\Delta t & = & h^{-1} \tau_{100} \ {\times} \nonumber \\
~ & ~ & \left [ \frac{1}{2} R^2 - R R_s \cos{(\vartheta - \vartheta_s)} + 
\frac{1}{2} R_s^2 - \psi(R, \vartheta) \right ] 
\label{eq: timedelaygen}
\end{eqnarray}
where $(R, \vartheta)$ is the image position, $(R_s, \vartheta_s)$ the unknown source position and we have defined\,:

\begin{equation}
\tau_{100} = \left(\frac{D_{l} D_{s}}{D_{ls}}\right) \frac{(1 + z_l)}{c} \ .
\label{eq: taucento}
\end{equation}
According to the Fermat principle, the images lie at the minima of $\Delta t$, so that the lens equations may be simply obtained by minimizing $\Delta t$. Inserting Eq.(\ref{eq: psirtm}) into Eq.(\ref{eq: timedelaygen}) and differentiating, we get\,:

\begin{equation}
\xi - \xi_s \cos{(\vartheta - \vartheta_s)} = \frac{\alpha_v}{R_v} \times \frac{{\cal{F}}(\xi, x_p)}{{\cal{F}}(1, x_p)}
\label{eq: lenseqa} 
\end{equation}

\begin{equation}
\xi_s \sin{(\vartheta - \vartheta_s)} = 0 \ .
\label{eq: lenseqb}
\end{equation} 
with $\xi_s = R_s/R_v$. Eq.(\ref{eq: lenseqb}) has two solutions. The first is $\xi_s = 0$, i.e. lens, source and observer are perfectly aligned. The only image is the Einstein ring that we will discuss in much detail in the next section. The second solution is obtained for $\sin{\vartheta - \vartheta_s} = 0 \iff \vartheta = \vartheta_s + m \pi$ with $m= 0, 1$. In this case, we get two images\footnote{Note that the theorem of the odd number of images (\cite{SEF,Straumann}) does not apply here because the mass distribution is singular in the origin.} symmetrically placed with respect to the lens centre. The radial coordinate $\xi_1$ of the first image (i.e., the one with $\vartheta = \vartheta_s$) is obtained by solving Eq.(\ref{eq: lenseqa}) with $\cos{(\vartheta - \vartheta_s)} = 1$, while the second has a distance $\xi_2$ from the lens centre obtained by solving Eq.(\ref{eq: lenseqa}) with $\cos{(\vartheta - \vartheta_s)} = -1$.  These equations may be solved numerically provided that the cluster parameters have been fixed. Up to now, there is only one multiple image system in which the lens is a cluster galaxy, namely the recently discovered SDSS\,J1004+4112 (\cite{SDSSlens1,SDSSlens2}), while for all the other multiply imaged quasar the lens is a galaxy (see, e.g., the CASTLES web page, \cite{CASTLES}). Since the RTM model has not been tested on galactic scale (given the mass range probed by the simulations employed by Rasia et al. 2003), we prefer to not discuss further the formation of multiple images.

\section{Critical curves}

The most spectacular effect of lensing by a galaxy cluster is the formation of giant arcs (see, e.g., \cite{Kneib96} for the textbook example of A2218). These are very luminous and highly distorted images of a source galaxy whose position is near one of the critical curves of the lensing potential. The position of the arcs in a given lensing system allows to strongly constrain the cluster mass distribution and can also be used to determine the cosmological parameters (\cite{Mauro02}). Moreover, arc statistics is a promising and efficient tool to discriminate among different cosmological models and theories of structure formation. It is thus quite interesting to investigate the number (and the type) of arcs the RTM model may form.

\begin{figure}
\centering
\resizebox{8.5cm}{!}{\includegraphics{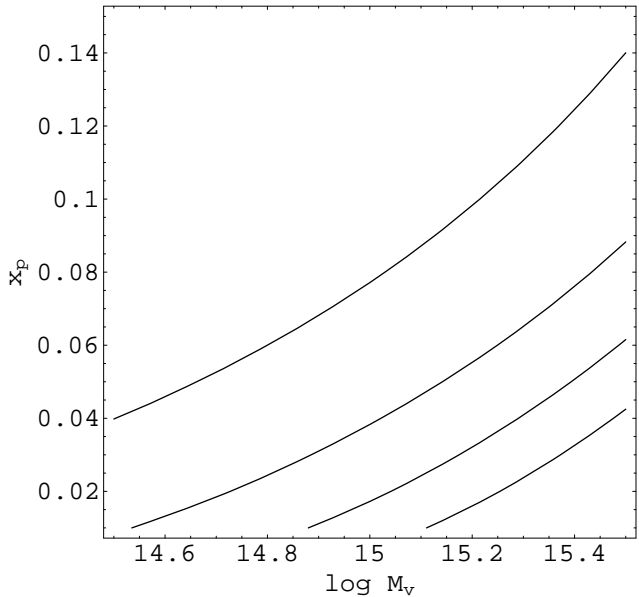}}
\hfill
\caption{Contours of equal Einstein radius $R_E$ in the $(\log{M_v}, x_p)$ plane. $R_E$ ranges from 5 (the uppermost curve) to 35\,arcsec (the lowermost one) in steps of 10\,arcsec.}
\label{fig: reconts}
\end{figure}

To this aim, let us first remember the expression for the magnification $\mu$ of a source due to the lensing effect. It is (\cite{SEF})\,:

\begin{equation}
\mu = \frac{1}{\det{A}} = \frac{1}{(1 - \psi_{xx}) (1 - \psi_{yy}) - \psi_{xy}^2} = \frac{1}{\lambda_r \lambda_t} 
\label{eq: defmu}
\end{equation}
where $A$ is the amplification matrix (that is the jacobian matrix of the lens mapping) and $(\lambda_r, \lambda_t)$ for a circularly symmetric model are given as\,:

\begin{equation}
\lambda_r = 1 - \frac{d\alpha}{dR} \ ,
\label{eq: deflambdar}
\end{equation}

\begin{equation}
\lambda_t = 1 - \frac{\alpha}{R} \ .
\label{eq: deflambdat}
\end{equation} 
Inserting Eq.(\ref{eq: alphartm}) into Eqs.(\ref{eq: deflambdar}), (\ref{eq: deflambdat}), we get the corresponding quantities for the RTM model that we do not explicitly report here for sake of shortness. The critical curves are the loci where $\det{A} = 0$. This condition is satisfied by imposing $\lambda_r = 0$ or $\lambda_t = 0$. The second equation implicitly defines the tangential critical curves. It is easy to show that\,:

\begin{equation}
\lambda_t = 0 \rightarrow \xi_E = \frac{\alpha_v}{R_v} \times \frac{{\cal{F}}(\xi_E, x_p)}{{\cal{F}}(1, x_p)} 
\label{eq: findxie}
\end{equation}
with $\xi_E \equiv R_E/R_v$ and $R_E$ the Einstein radius. Eq.(\ref{eq: findxie}) may be solved numerically for fixed values of the cluster model parameters. Fig.\,\ref{fig: reconts} shows the contours of equal $R_E$ in the $(\log{M_v}, x_p)$ plane\footnote{Unless otherwise stated, $\log{x}$ is the logarithm base 10.} with higher values of $R_E$ corresponding to lower curves in the plot. We have only considered clusters with $\log{M_v} \in (14.5, 15.5)$ because this is (approximately) the mass range probed in Rasia et al. (2004), while we will (usually but not always) consider $x_p \in (0.01/7.13, 2/7.13)$ since Rasia et al. (2004) states that the average value of $\langle c_{RTM} \rangle = \langle 1/x_p \rangle \simeq 7.13$ over their cluster sample. 

\begin{figure}
\centering
\resizebox{8.5cm}{!}{\includegraphics{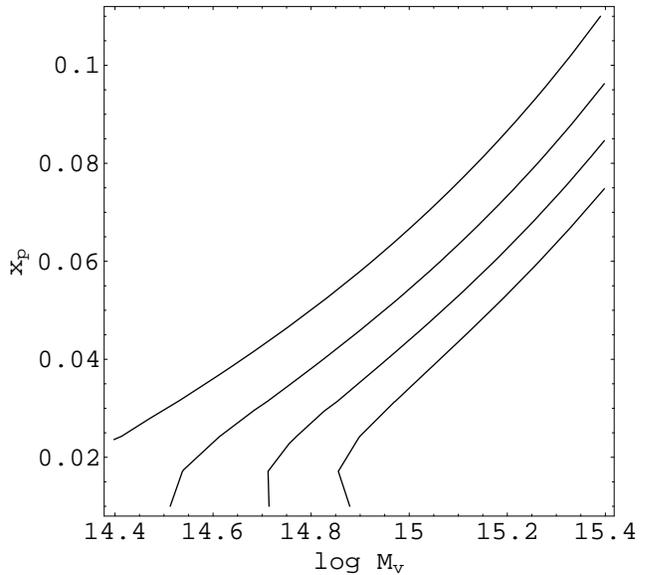}}
\hfill
\caption{Contours of equal radial critical curve distance $R_{rad}$ from the cluster centre in the $(\log{M_v}, x_p)$ plane. $R_{rad}$ ranges from 2.5 (the uppermost curve) to 5.5\,arcsec (the lowermost one) in steps of 1\,arcsec.}
\label{fig: lr}
\end{figure}

Fig.\,\ref{fig: reconts} shows that highly concentrated clusters (i.e., with small values of $x_p$) give rise to tangential critical curves that are more distant from the cluster centre. Thus, one could qualitatively conclude that only RTM models with lower values of $x_p$ are able to produce tangential arcs. However, one should also take into account that, as expected, for a fixed $x_p$ the Einstein radius increases with $M_v$. Therefore, RTM models with high values of $x_p$ could still produce tangential arcs with large radii provided that the mass is large enough and that it is possible to extrapolate the RTM model outside the mass range probed by the simulations.  

As said above, radial critical curves are implicitly defined by the condition $\lambda_r = 0$. Inserting Eq.(\ref{eq: sigmartm}) and (\ref{eq: alphartm}) into Eq.(\ref{eq: deflambdar}), we get $R_{rad}$, the radial critical curve distance from the cluster centre. To investigate how $R_{rad}$ depends on the model parameters, we plot in Fig.\,\ref{fig: lr} the contours of equal $R_{rad}$ in the $(\log{M_v}, x_p)$ plane. The behaviour of $R_{rad}$ with the $\log{M_v}$ is qualitatively similar to that of the Einstein radius $R_E$, but the opposite holds for the dependence on $x_p$. For a given value of $\log{M_v}$, $R_{rad}$ increases with $x_p$, while $R_{E}$ decreases. As a result, highly concentrated RTM models (i.e. models with small $x_p$ and henceforth high $c_{RTM}$) give rise to tangential arcs situated at large distances from the cluster centre, but the radial arc lies in the very inner regions of the cluster.

As a simple application of these results, we consider the case of the real cluster lens MS2137-23 in which both a tangential and a radial arc have been observed with $R_E = 15.35$\,arcsec and $R_{rad} = 4.5$\,arcsec (Sand et al. 2002, 2004). An easy way to find out the values of the RTM model parameters able to fit the arc positions in MS2137-23 is to draw the contour levels in the plane $(\log{M_v}, x_p)$ for $R_E$ and $R_{rad}$ equal to the values quoted above and look for the intersection point between these two curves. It turns out that the best fit parameters are $(\log{M_v}, x_p) = (14.7, 0.019)$, i.e. $M_v = 5.0 \times 10^{14} \ M_{\odot}$ and $c_{RTM} \simeq 53$. As a further test, we also consider the case of the real cluster lens RXJ1133 at redshift $z_L = 0.394$ (\cite{STSE03}). Both a radial and a tangential arc are observed with $z_S = 1.544$ and $(R_{rad}, R_E) = (3.2, 10.9) \ arcsec$ that may be obtained by describing the cluster with a RTM model with best fit parameters $(\log{M_v}, x_p) = (14.5, 0.018)$, i.e. $M_v = 3.2 \times 10^{14} \ {\rm M_{\odot}}$ and $c_{RTM} \simeq 56$. The values of the concentration are quite high if compared to $\langle c_{RTM} \rangle \simeq 7.13$ found by Rasia et al. (2004) over their sample of simulated clusters. However, this could not be considered an evidence against the RTM model. Actually, we have only considered the spherical case, while it is well known that also a small cluster ellipticity changes significantly the position of the critical curves (see, e.g., \cite{BartMen03}). Moreover, one should also take into account the impact on the critical curves of the galaxy lying at the centre of the cluster gravitational potential and of other eventual substructures. That is why we do not speculate further on the high $c_{RTM}$ values needed to reproduce the arcs positions in MS2137-23 and RXJ1133, while a detailed comparison with observations will be presented elsewhere. 

\section{Adding a bright cluster galaxy}

Some recent studies have highlighted the importance of considering the brightest cluster galaxy (hereafter BCG) when investigating the lensing properties of a cluster (\cite{MBM03,STSE03}). It is thus interesting to study how the critical curves of the RTM model are affected by the addition of a BCG. To this aim, we place the galaxy exactly at the centre of the cluster and model it using the Hernquist profile whose mass density is (\cite{H90})\,:

\begin{equation}
\rho(r) = \frac{\rho_s}{r/r_s \ (1 + r/r_s)^3}
\label{eq: rhohern}
\end{equation}
with $\rho_s$ a characteristic density and $r_s$ a scale radius. The Hernquist profile has the notable property that its projected density well approximates the $R^{1/4}$ law (\cite{deV48}) provided that the effective radius is related to the scale radius of the Hernquist model by the relation\,: $R_e \simeq 1.81 r_s$. The model is fully characterized by two parameters that we choose to be $R_e$ and the total mass $M_h$ given by\,:

\begin{equation}
M_h = 2 \pi r_s^3 \ \rho_s \ .
\label{eq: masshern}
\end{equation}
Assuming spherical symmetry, the deflection angle of the galaxy is (\cite{K01})\,:

\begin{equation}
\alpha_H(\sigma) = 2 \kappa_s r_s \ \frac{\sigma [1 - {\cal{H}}(\sigma)]}{\sigma^2 - 1} 
\label{eq: alphahern}
\end{equation}
with $\sigma = R/r_s$ and $\kappa_s = \rho_s r_s/\Sigma_{crit}$ and we have defined\,:

\begin{equation}
{\cal{H}}(\sigma) = \cases{
  {1 \over \sqrt{\sigma^2-1}}\,\mbox{tan}^{-1} \sqrt{ \sigma^2-1 } & $(\sigma>1)$ \cr
  {1 \over \sqrt{1-\sigma^2}}\,\mbox{tanh}^{-1}\sqrt{ 1-\sigma^2 } & $(\sigma<1)$ \cr
  1                                                      & $(\sigma=1)$ \ . \cr }
\label{eq: defacca}
\end{equation}
Since the standard theory of lensing is developed in the weak field limit, the total deflection angle is simply the sum of the contributions from the cluster and the BCG. Hence, the total magnification may be written as\,:

\begin{figure}
\centering
\resizebox{8.5cm}{!}{\includegraphics{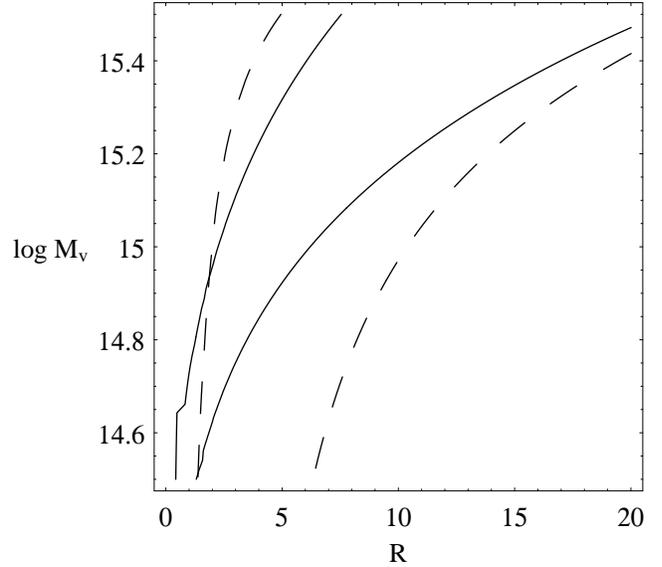}}
\hfill
\caption{Zero level curves of the denominator in Eq.(\ref{eq: mutot}) in the plane $(R, \log{M_v})$ for a RTM cluster with $x_p = 0.5/7.13$. The solid line refers to the case with no BCG, while the dashed one has been obtained adding a BCG with $(M_h, R_e) = (5 \times 10^{12} \ {\rm M_{\odot}}, 24.80 \ {\rm kpc})$. For a given value of $M_v$, the intersections of the horizontal line $\log{M_v} = const$ with the curves plotted are the values of $R_{rad}$ and $R_E$ respectively.}
\label{fig: detA-RTM+Hern}
\end{figure}

\begin{equation}
\mu = \frac{1}{(\lambda_{r}^{H} + \lambda_{r}^{RTM} - 1) (\lambda_{t}^{H} + \lambda_{t}^{RTM} - 1)}
\label{eq: mutot}
\end{equation}
with $\lambda_r$ and $\lambda_t$ given by Eqs.(\ref{eq: deflambdar}) and (\ref{eq: deflambdat}) respectively and quantities with the superscript {\it ``H''} ({\it ``RTM''}) refers to the Hernquist (RTM) model. The radial and tangential arcs radii $R_{rad}$ and $R_E$ are defined as those radii vanishing the first and the second term respectively of the denominator in Eq.(\ref{eq: mutot}). 

\begin{figure}
\centering
\resizebox{8.5cm}{!}{\includegraphics{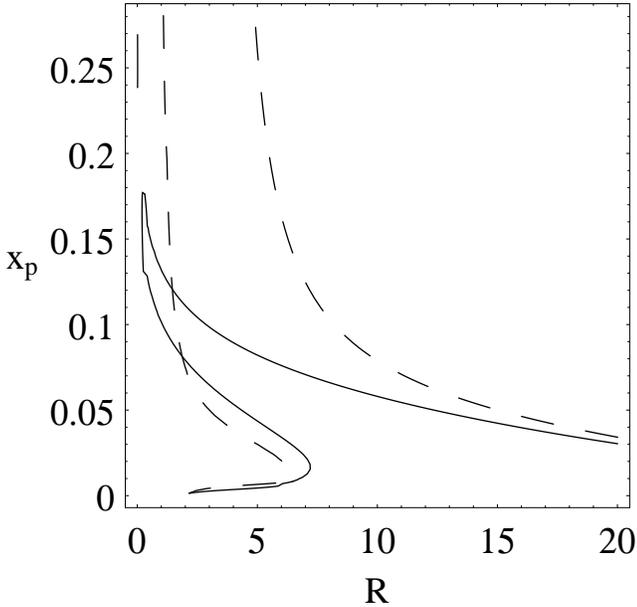}}
\hfill
\caption{Same as Fig.\,\ref{fig: detA-RTM+Hern} but now the value of $M_v$ is fixed as $1.125 \times 10^{15} \ {\rm M_{\odot}}$ and $x_p$ is changing. Solid line refers to the case with no BCG, while dashed one to the case with a BCG with parameters fixed as before.}
\label{fig: detA-RTM+Hern-bis}
\end{figure}

To study where the critical curves form, we plot in Fig.\,\ref{fig: detA-RTM+Hern} the loci in the $(R, \log{M_v})$ plane where the total magnification formally diverges having arbitrarily fixed $x_p = 0.5/7.13$ and $R_e = 24.80$\,kpc as for the BCG in MS2137-23 (\cite{STSE03}). The solid line refers to the case with no BCG, while the dashed one shows how the curves are modified by the addition of a BCG with total mass\footnote{Note that the value chosen for the mass of the BCG, even if high, is not unrealistic. For instance, the estimated total mass of the Milky Way is $1.9_{-1.7}^{+3.6} \times 10^{12} \ {\rm M_{\odot}}$ (\cite{WE99}) which is in the mass range we have adopted for the BCG.} $M_h = 5 \times 10^{12} \ {\rm M_{\odot}}$. The number of critical curves is still two, but their position is affected by the presence of BCG with the distance between them increased with respect to the case with no BCG. In particular, while $R_{rad}$ is slightly smaller or higher depending on the ratio between the mass of galaxy and that of the cluster, the Einstein radius $R_E$ significantly increases. This is expected since $R_E$ is proportional to the total mass within the tangential critical curve so that, adding the BCG mass, $R_E$, gets obviously higher. 

Fig.\,\ref{fig: detA-RTM+Hern-bis} is similar to Fig.\,\ref{fig: detA-RTM+Hern}, but now the cluster mass is set as $M_v = 1.125 \times 10^{15} \ {\rm M_{\odot}}$ and we let $x_p$ changing, while the BCG parameters are fixed as before. The increase of $R_E$ is still visible, but what is most important to note is the possibility to have radial and critical curves with higher values of $x_p$. From Fig.\,\ref{fig: detA-RTM+Hern-bis}, one sees that, when the BCG is absent, RTM models with $x_p > 0.18$ are unable to produce radial arcs (i.e. it is $R_{rad} = 0$), while the radial critical curve appears when the BCG is taken into account in the total lensing potential even for less concentrated (i.e. with larger $x_p$) clusters. 

\begin{figure}
\centering
\resizebox{8.5cm}{!}{\includegraphics{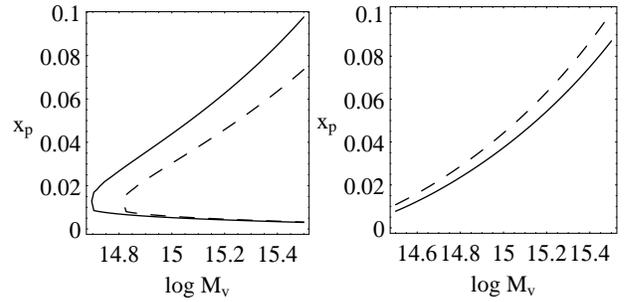}}
\hfill
\caption{Constraints in the $(\log{M_v}, x_p)$ plane imposing $R_{rad} = 4.5 \ {\rm arcsec}$ (left) or $R_E = 15.35 \ {\rm arcsec}$ (right) for the RTM model with (dashed line) and without (solid line) taking into account the presence of a BCG. The galaxy parameters are set as $(M_h, R_e) = (5.0 \times 10^{12} \ {\rm M_{\odot}}, 24.80 \ {\rm kpc})$.}
\label{fig: arcsconstr-Hern}
\end{figure}

Finally, we investigate qualitatively how the constraints on the RTM model parameters are changed by the presence of a BCG. To this aim, we plot in Fig.\,\ref{fig: arcsconstr-Hern} the constraints imposed by the presence of a radial arc at $R_{rad} =  4.5 \ {\rm arcsec}$ or a tangential arc at $R_E = 15.35 \ {\rm arcsec}$ as observed for the real cluster lens MS2137-23 (Sand et al. 2002, 2004). The results in the right panel are easy to explain qualitatively. Adding a BCG pushes up the curve in the $(\log{M_v}, x_p)$ plane so that a tangential arc at a given distance may be produced by less concentrated and less massive clusters with respect to the case with no BCG. As yet noted, $R_E$ is proportional to the total projected mass within $R_E$ itself. Since now the BCG provides part of this mass, less mass has to be contributed by the cluster and thus less massive and concentrated models are needed to obtain a given value of $R_E$. Note, however, that the deviations from the case with no BCG are quite small as expected given the high mass ratio between the cluster and the galaxy. On the other hand, the left panel shows that adding a BCG requires more concentrated and massive clusters to produce a radial arc at a given $R_{rad}$ with respect to the case with no BCG. It is also worth noting that the constraints from the position of the radial arc are more sensitive to the presence (or absence) of the BCG (see, e.g., the distance between the solid and dashed line in the left panel compared to the same in the right one). This is expected since the radial critical curve is innermost and thus probes a range that is more sensitive to the inner structure of the cluster where the BCG plays a more significant role.

\section{The impact of the shear}

Up to now, we have investigated the lensing properties of the RTM model (with and without a central BCG) assuming spherical symmetry of the mass distribution. However, real clusters are moderately elliptical and it is well known that even small ellipticities may alter significantly the lensing properties of a given model. In particular, taking into account deviations from spherical symmetry is very important when trying to extract constraints on the model parameters from the position of the lensed arcs in real systems (as clearly demonstrated, for instance, in \cite{BartMen03,DalKee03}). On the other hand, even when assuming spherical symmetry, it is important to take into account also the effect of substructures in the cluster mass distribution and possible tidal deformations due to nearby clusters. Finally, the large scale structure as a whole could also have a not negligible effect (\cite{KKS97}). 

To the lowest order, all these effects may be mimicked by adding an external shear to the lensing potential which is now written as\,:

\begin{figure}
\centering
\resizebox{8.5cm}{!}{\includegraphics{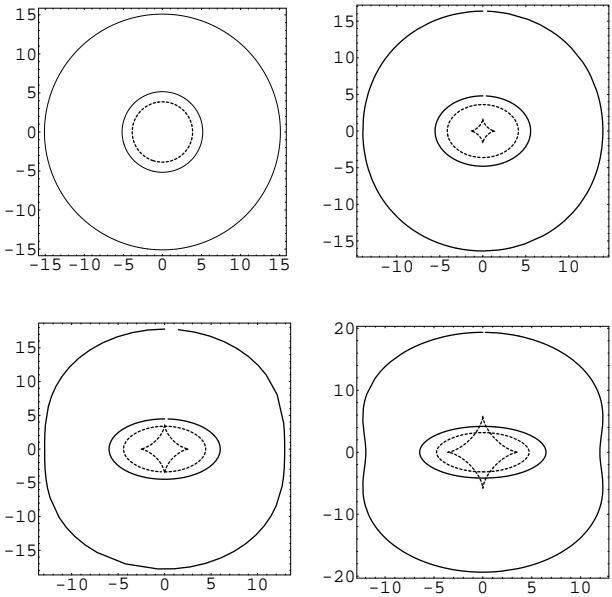}}
\hfill
\caption{Critical curves (solid) and caustics (dashed) for the RTM model with the addition of an external shear. Upper panels are for $\gamma = 0.0$ (left) and $\gamma = 0.05$ (right), while lower panels refer to $\gamma = 0.10$ (left) and $\gamma = 0.15$ (right). The shear position angle is set at $\theta_{\gamma} = 0$, while the cluster parameters are $(M_v, x_p) = (1.125 \times 10^{15} \ {\rm M_{\odot}}, 0.03)$.}
\label{fig: shearplot}
\end{figure}

\begin{equation}
\psi(R, \vartheta) = \psi_{RTM}(R) - \frac{1}{2} \gamma R^2 \cos{2(\vartheta - \vartheta_{\gamma})}
\label{eq: psitot}
\end{equation}
with $\psi_{RTM}$ given by Eq.(\ref{eq: psirtm}) and $(\gamma, \vartheta_{\gamma})$ the shear strength and position angle. Without loss of generality we assume that the shear is oriented along the major axis so that it is $\vartheta_{\gamma} = 0$. Fig.\,\ref{fig: shearplot} shows the critical curves and the caustics for the lensing potential in Eq.(\ref{eq: psitot}). To be quantitative, we also report in Table\,1 the quantity\,:

\begin{displaymath}
\Delta \xi_i = 100 \times \frac{\xi_i(\gamma) - \xi_i(\gamma = 0)}{\xi_i(\gamma = 0)}
\end{displaymath}
with $\xi_i$ the length of the radial or tangential critical curve along the $i$\,-\,axis (with $i = x, y$).

In the case with no shear (upper left panel), the two critical curves are spherical, while the tangential caustic is the origin and the radial one is a circle. The effect of the shear is to deform the critical curves into ellipses, while the inner caustic (corresponding to the tangential critical curve) takes a diamond shape and the external one becomes elliptical. In particular, the radial critical curve is more and more elongated along the major axis\footnote{This is a consequence of having fixed $\vartheta_{\gamma} = 0$. In general, the ellipse is elongated along the direction individuated by $\vartheta_{\gamma}$.} as the shear strength increases. Quantitatively, this could be seen in Table\,1 where $\Delta \xi_x$ is positive and increases with $\gamma$, while $\Delta \xi_y$ is negative and different from $\Delta \xi_x$. This is what corresponds graphically to an ellipse more and more elongated along the major axis as $\gamma$ increases. On the other hand, the tangential critical curve may also deviate from the elliptical symmetry taking a dumbell shape as $\gamma$ gets higher. As a consequence, the higher is $\gamma$, the higher is the the radial arc distance (measured along the major axis), $R_{rad}$, from the cluster centre for fixed values of the RTM model parameters. A similar result holds also for $R_E$. It is worth noting that these effects are qualitatively the same whatever is the value of $x_p$, but are more pronounced (i.e. $\Delta \xi_i$ is larger) for lower values of $x_p$ (i.e. higher concentrations) as can be seen from Table\,1.

\begin{table}
\begin{center}
\begin{tabular}{|c|c|c|c|c|}
\hline
$\gamma$ & \multicolumn{2}{|c|}{Tangential} & \multicolumn{2}{|c|}{Radial} \\
\hline 
~ & $\Delta \xi_x$ & $\Delta \xi_y$ & $\Delta \xi_x$ & $\Delta \xi_y$ \\ 
\hline
\hline
0.05 & -8 & 8 & 8 & -5 \\
0.10 & -14 & 18 & 16 & -13 \\
0.15 & -21 & 28 & 25 & -19 \\
\hline
0.05 & -12 & 13 & 13 & -12 \\
0.10 & -22 & 29 & 28 & -22 \\
0.15 & -31 & 46 & 45 & -31 \\
\hline
\end{tabular}
\end{center}
\caption{The impact of the shear on the length along the major an minor axes of the tangential and radial critical curves. The upper half of the table refers to RTM model with $x_p = 1/\langle c_{RTM} \rangle$, while for the lower half it is $x_p = 0.5 \langle c_{RTM} \rangle$. In both cases, the virial mass is set to $M_v = 1.125 \times 10^{15} \ {\rm M_{\odot}}$.}
\end{table}

Note that these results are in qualitative agreement with the approximate analytical treatment presented in Bartelmann \& Meneghetti (2004). Actually, these authors used a different approach deforming the lens model so that the isocountour lines of the lensing potential are ellipses with ellipticity $\varepsilon$.  Elliptical deformation of the lensing potential leads to dumbell shaped surface mass distribution for values of $\varepsilon > 0.2$. Even if clusters are highly structured, similar mass models are quite unrealistic so that we have preferred not to follow this approach. On the other hand, it is possible to show that an elliptical potential $\psi(x^2 + y^2/q^2)$ with an on axis shear and axial ratio $q$ produces the same image configuration as a pure elliptical potential with axis ratio $q' = q \sqrt{(1-\gamma)/(1+\gamma)}$ without shear (\cite{W96}). In our case, this means that using the lensing potential given by Eq.(\ref{eq: psitot}) is equivalent to deform the RTM model such that the lensing isopotential contours are ellipses with axis ratio $q' = \sqrt{(1-\gamma)/(1+\gamma)}$. This shows the complete equivalence among our approach and that of Bartelmann \& Meneghetti (2004). 

\section{Comparison with the NFW model}

It is interesting to compare the lensing properties of the RTM model with those of the model proposed by Navarro, Frenck \& White (1997, hereafter NFW) and mostly used in literature. Using the same normalization as in Rasia et al. (2004), the density profile of the NFW model is\,:

\begin{equation}
\rho = \frac{\rho_{0,NFW} \rho_b}{(x/x_s) (1 + x/x_s)^2} 
\label{eq: rhonfw}
\end{equation}
with $x = r/R_v$, $x_s \equiv 1/c_{NFW}$, $c_{NFW}$ the concentration of the NFW model and\,:

\begin{equation}
\rho_{0,NFW} = \frac{(1 - f_b) \Delta_v}{3 \left [ \ln{(1 + c_{NFW})} - c_{NFW}/(1 + c_{NFW}) \right ]} \ .
\label{eq: rhoznfw}
\end{equation}
The deflection angle for the NFW model may be conveniently written as (\cite{B96,K01})\,:

\begin{figure}
\centering
\resizebox{8.5cm}{!}{\includegraphics{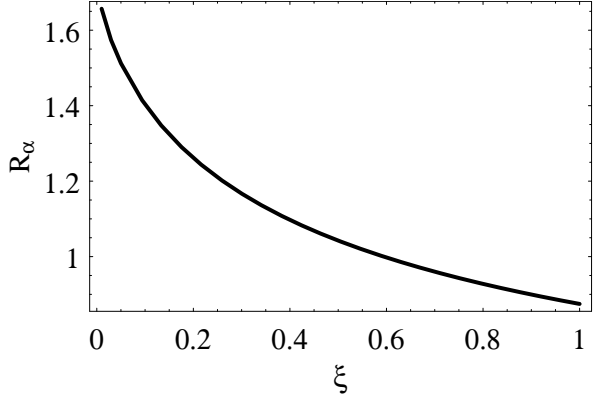}}
\hfill
\caption{${\cal{R}}_{\alpha}$ vs $\xi$ for the average values of the concentration parameters of the NFW and RTM model.}
\label{fig: ratiovsxi}
\end{figure}

\begin{equation}
\alpha_{NFW}(\xi) = \frac{\alpha_{v}^{NFW}}{\xi} \times 
\frac{\ln{(c_{NFW} \xi/2)} + {\cal{H}}(c_{NFW} \xi)}{\ln{(c_{NFW}/2)} + {\cal{H}}(c_{NFW})} 
\label{eq: alphanfw}
\end{equation}
having defined\,:

\begin{eqnarray}
\alpha_{v}^{NFW} & = & \frac{4 (1 - f_b) \Delta_v \rho_b}{3 \Sigma_{crit}} \times R_v \times \nonumber \\ 
~ & ~ & \frac{\ln{(c_{NFW}/2)} + {\cal{H}}(c_{NFW})}{\ln{(1 + c_{NFW})} - c_{NFW}/(1 + c_{NFW})} 
\label{eq: defavnfw}
\end{eqnarray}
with $R_v$ expressed in {\it arcsec}. 

\begin{figure}
\centering
\resizebox{8.5cm}{!}{\includegraphics{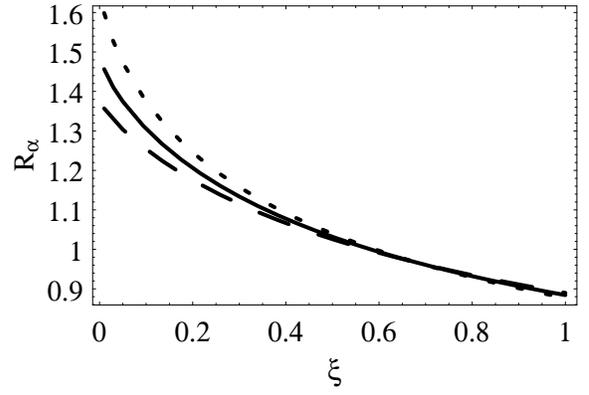}}
\hfill
\caption{${\cal{R}}_{\alpha}$ vs $\xi$ setting the NFW concentration to the value predicted by Eq.(\ref{eq: cvsmv}) and $c_{RTM} = 7.13$ for three values of the virial mass, namely $M_v = 5 \times 10^{14}$ (short dashed), $M_v = 7.5 \times 10^{14}$ (solid), $M_v = 10^{15}$ (long dashed).}   
\label{fig: ratiomass}
\end{figure}

Using Eqs.(\ref{eq: alphartm}), (\ref{eq: alphav}), (\ref{eq: alphanfw}) and (\ref{eq: defavnfw}) and the definition of virial radius, we get the following expression for the ratio between the deflection angle of the NFW and RTM model\,:

\begin{eqnarray}
{\cal{R}}_{\alpha} \equiv \frac{\alpha_{NFW}}{\alpha_{RTM}} & = & \frac{2 \sqrt{\pi/8}}{\xi} \left ( \frac{M_{v}^{NFW}}{M_v} \right )^{1/3} \times \nonumber \\
~ & ~ & \frac{(1 + 2 x_p)/\sqrt{1 + x_p} - 2 \sqrt{x_p}}{\ln{(1 + c_{NFW})} - c_{NFW}/(1 + c_{NFW})} \times \nonumber \\
~ & ~ & \frac{\ln{(c_{NFW} \xi/2)} + {\cal{H}}(c_{NFW} \xi)}{{\cal{F}}(\xi, x_p)} \ .
\label{eq: ratio}
\end{eqnarray}
It is worth noting that the ratio does not depend on the redshift of lens and source as it is expected since it is related to the different density profiles of the two models which is, of course, the same at all redshifts. Moreover, to better compare the lensing properties of the two models, it is meaningful to assume that the virial mass is the same so that, at a given radius $\xi$,  ${\cal{R}}_{\alpha}$ only depends on the concentrations $c_{RTM}$ and $c_{NFW}$ of the two models. Fig.\,\ref{fig: ratiovsxi} shows ${\cal{R}}_{\alpha}(\xi)$ setting $c_{RTM} = 7.13$ and $c_{NFW} = 6.8$ as determined by Rasia et al. (2004) averaging over their simulated clusters sample. The deflection angle of the NFW model turns out to be higher than that of the RTM model until $\xi < 0.6$, while ${\cal{R}}_{\alpha} < 1$ for light rays impacting in the outer region of the halo. This result is simply related to the different mass profile of the two models with $M_{NFW}(\xi)/M_{RTM}(\xi)$ being larger than 1 in the inner regions for this choice of $(c_{NFW}, c_{RTM})$.

According to some authors, the NFW model is a one parameter model since it is possible to relate the concentration $c_{NFW}$ to the virial mass even if this relation has a quite large scatter. Following Bullock et al. (2001), we adopt\,:

\begin{equation}
c_{NFW} = 15 - 3.3 \log{\frac{M_v}{10^{12} h^{-1} \ {\rm M_{\odot}}}} 
\label{eq: cvsmv}
\end{equation} 
and plot, in Fig.\,\ref{fig: ratiomass}, ${\cal{R}}_{\alpha}$ for three different values of the virial mass. The qualitative behaviour is the same, but it is more pronounced for lower mass models corresponding, according to Eq.(\ref{eq: cvsmv}), to lower concentrations.  

\begin{figure}
\centering
\resizebox{8.5cm}{!}{\includegraphics{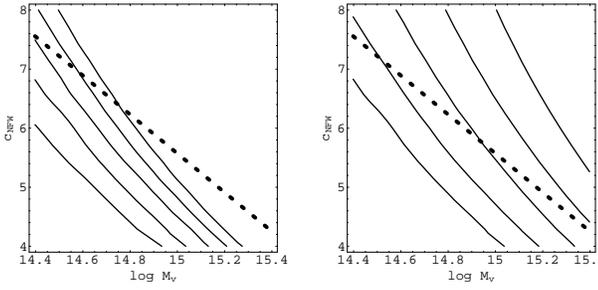}}
\hfill
\caption{Contour plot in the $(\log{M_v}, c_{NFW})$ plane for the Einstein radius of the NFW model. {\it Left panel}\,: contours corresponding to values of $R_E$ for the RTM model with $x_p = 1/\langle c_{RTM} \rangle$ and $M_v$ from 7.5 to $17.5 \times 10^{14} \ {\rm M_{\odot}}$ (from left to right) in steps of $2.5 \times 10^{14} \ {\rm M_{\odot}}$. {\it Right panel}\,: contours corresponding to values of $R_E$ for the RTM model with $M_v = 10^{15}$ and $x_p$ from 0.06 to $0.14$ (from right to left) in steps of 0.02. The dashed line refers to Eq.(\ref{eq: cvsmv}).}
\label{fig: ratiore}
\end{figure}

The NFW model is used in most of the studies of the structure of dark matter haloes in galaxy clusters. It is therefore interesting to investigate what is the error induced by using the NFW model to describe a cluster that is actually better described by the RTM model. As a straightforward example, we consider the estimate of the virial mass from the size of the Einstein radius. To this aim, let us look at Fig.\,\ref{fig: ratiore} where we plot the contour level curves in the $(\log{M_v}, c_{NFW})$ plane for the Einstein radius of the NFW model corresponding to values of $R_E$ evaluated for RTM models. Let us consider, for instance, the RTM model with $(M_v, x_p) = (1.5 \times 10^{15} \ {\rm M_{\odot}}, 1/7.13)$ for which one obtains the fourth line (from left) in the left panel of Fig.\,\ref{fig: ratiore}. If one assumes that Eq.(\ref{eq: cvsmv}) holds, then one should estimate the NFW model parameters from the intersection point of the dashed curve with the fourth line thus grossly underestimating the mass. Actually, even if one does not use Eq.(\ref{eq: cvsmv}), $M_v$ turns out to be underestimated for the values of $c_{NFW}$ in the plot. To get the correct value of $M_v$, one should select a value of $c_{RTM}$ that is unrealistically low for a galaxy cluster. We thus conclude that using the NFW model to study a cluster that is intrinsically described by the RTM model leads to underestimate the virial mass by an amount that depends on the concentration of both the NFW and the RTM model.

In principle, one should compare the lensing properties of the NFW and of the RTM models by also including the effects of the brightest cluster galaxy and of the shear. However, this should increase the number of parameters to eight\,: two for the NFW model, two for the RTM model, the BCG mass and scale radius and the shear strength and orientation.  It is likely that some degeneracies could occur among parameters rendering a comparison between the NFW and the RTM model meaningless in this case so that we prefer to not perform this test.

\section{The thermal Sunyaev\,-\,Zel'dovich effect}

At the beginning of Seventies, Sunyaev and Zel'dovich (1970, 1972) suggested that the cosmic microwave background radiation (CMBR) can be scattered by the trapped hot intracluster electrons giving rise to a measurable distortion of its spectrum. This (inverse) Compton scattering (now referred to as {\it Sunyaev\,-\,Zel'dovich effect}, hereafter SZE) has been recognized in the last two decades as an important tool for cosmological and astrophysical studies (\cite{Bir99}). More recently, the SZE has been used to investigate several physical properties of the gas with much attention devoted to the geometry of its density profile as well as its thermodynamical status. In particular, it has been shown that the {\it canonical} hypotheses of spherical symmetry and isothermal temperature profile may induce errors (of order up to $30\%$) on the estimated values of different parameters such as the Hubble constant (\cite{RSM97,jetzer1}).

The physics of the SZE is quite simple to understand. A gas of electrons in hydrostatic equilibrium within the gravitational potential of a cluster will have a temperature $T_e \simeq G M m_p/(2 k_B R_{eff}) \sim {\rm few \ keV}$, with $M$ and $R_{eff}$ typical values of the total mass and size of a cluster. At this temperature, the thermal emission in X\,-\,ray is composed of thermal bremsstrahlung and line radiation processes. Electrons in the intracluster gas are not only scattered by ions, but can themselves scatter photon of the CMBR giving rise, in average, to a slight change in the photon energy. Because of this inverse Thomson scattering, an overall change in brightness of the CMBR is observed. As a consequence, the SZE is localized and visible in and towards clusters of galaxies having an X\,-\,ray emission strong enough to be detectable.

In the non relativistic limit, the scattering process can be described by the Kompaneets equation

\begin{equation}
\frac{\partial n}{\partial y} = \frac{1}{p_e^2}
\frac{\partial}{\partial p_e} \left [ p_e^4 \left ( \frac{\partial
n}{\partial p_e} + n + n^2 \right ) \right ] \label{eq:sze1}
\end{equation}
which describes the change in the occupation number of photons $n(\nu)$. In Eq.(\ref{eq:sze1}), we have introduced the dimensionless variable\footnote{Here, we denote with the underscript {\it ``e''} quantities referring to electrons, while the same quantities without underscript refers to photons.} $p_e = h \nu_e/k_B T_e$, while $y$ is the so called Comptonization parameter defined as\,:

\begin{equation}
y \equiv \int{\frac{k_B T_e}{m_e c^2} n_e \sigma_T dl} \label{eq:
defycomp}
\end{equation}
being $\sigma_T$ the Thompson scattering cross section and $n_e$ the electrons number density and the integral is performed along the line of sight. Assuming that the photons distribution after the scattering is close to the equilibrium one, we have that\,:

\begin{equation}
\frac{\partial n}{\partial p_e}\gg n , n^2 \ ,
\end{equation}
so that Eq.(\ref{eq:sze1}) simplifies to

\begin{equation}
\frac{\partial n}{\partial y} = \frac{1}{p^2} \left( p^4
\frac{\partial n}{\partial p}\right) \ . \label{eq:sze3}
\end{equation}
having also replaced $p_e$ with $p$ because of the homogeneity. By solving this equation in the quasi equilibrium hypothesis, it is possible to obtain both the variation $\Delta n$ in the occupation number with respect to the equilibrium value $n_0$ and the corresponding shift in temperature $\Delta T$. It is\,:

\begin{equation}
\frac{\Delta n}{n_0} = \frac{y \ p \ e^p}{e^p -1} \left[ p \coth
\left( \frac{p}{2}\right) - 4 \right] \ , \label{eq:sze4}
\end{equation}

\begin{equation}
\frac{\Delta T}{T_0} = y\left[ p \, \mbox {coth}\,
\left(\frac{p}{2} \right) -4 \right] \equiv y \ g(p)\ , \label{eq:sze5}
\end{equation}
where $g(p)$ is the SZE frequency spectrum and we have considered that $T_e$ ($\sim 10^7\, K$) is much higher than the CMBR temperature $T_0 \simeq 2.7K$. In the limit of low frequencies, we get the useful approximated expression\,:

\begin{equation}
\frac{\Delta T}{T} = -2 y \ . \label{eq:sze5bis}
\end{equation}
Eq.(\ref{eq:sze5bis}) allows to evaluate the shift in temperature due to the SZE provided that the gas number density $n_e(r)$ and the temperature profiles $T_e(r)$ are given. For the RTM model, it is (\cite{RTM03})\,:

\begin{equation}
n_e(s) = \frac{\rho_{g,0} \ \rho_b}{\mu \ m_p} (s + x_p)^{-2.5} \
, \label{eq: nertm}
\end{equation}

\begin{equation}
T_e(s) =  \frac{T_0 T_v s^{0.016}}{\left ( s^4 + x_p^4 \right
)^{0.13}} \label{eq: tertm}
\end{equation}
where $s = r/R_v$, $\mu$ is the mean molecular weight, $m_p$ the proton mass, $T_v$ the virial temperature and $\rho_{g,0}$ a normalization density given by\,:

\begin{equation}
\rho_{g,0} = \displaystyle{ \frac{f_b \Delta_v} {3 \left [
\displaystyle{\frac{2 + 10 x_p + (40/3) x_p^2 + (16/3) x_p^3}{(1 +
x_p)^{2.5}}} - \displaystyle{\frac{16}{3} x_p^{1/2}} \right ]}}
\label{eq: rhozgas}
\end{equation}
Finally, in Eqs.(\ref{eq: nertm}) and (\ref{eq: tertm}), $x_p$ and $T_0$ are fitting parameters to be determined on a cluster by cluster basis. In particular, we set $(T_0, x_p) = (0.255, 10^{-0.51})$ as found by Rasia et al. (2004) for their set of simulated clusters.

In the following, we analyze the implications of the gas density and temperature profile of the RTM model on the SZE. Contrary to the lensing applications, we limit our analysis to the spherically symmetric case, without considering any ellipticity in the profiles. The reason for this choice is the fact that the gas profiles have been deduced by the mean of hydrodynamical (and not simply N\,-\,body) simulations, so that the hypothesis of axial symmetry for the density profile is not  enough to assure a {\it similar} elliptical temperature profile.

\subsection{The gas profile and the structure integral}

The radial dependence of cluster profiles is becoming a testing ground for models of structure formation and for our understanding of gas dynamics in galaxy clusters. Actually, the formation of structures is believed to be driven by some hierarchical development, which leads to the prediction of self similar scalings between systems of different masses and at different epochs. Moreover, the intracluster gas is generally assumed to be isothermal and in hydrostatical equilibrium. From the observational point of view, however, the situation is rather controversial and yet undeterminated\,: X\,-\,rays observations of poor clusters fall belove the self similar expectations, and even if the isothermal distribution is often a reasonable approximation to the actual observed clusters, some clusters show not isothermal distribution (\cite{jetzer}). It turns out that the emerging temperature profile is one where the temperature increases from the center to some characteristic radius, and then decreases again. The central temperature decrements has been much discussed in terms of cooling flows, while the outer temperature decrement seems to be confirmed both observationally and numerically. On the other hand, the temperature profile described in Eq.(\ref{eq: tertm}) reproduces quite well some of these observational features\,: it shows an isothermal core up to $0.2 \,R_v$, followed by a steep decrease that reaches a factor two lower around the virial radius; the density profiles are self\,-\,similar roughly $s > 0.06$, while the gas becomes flatter in the inner region. However, these non\,-\,trivial clusters profiles need to be observationally tested. The observations of the SZE, which are becoming increasingly accurate, can be used to probe these properties. Here we analyze the radial dependence of the SZE observables, deserving the comparison with the observational data in the X\,-\,ray and SZE domain to a forthcoming paper.

The temperature shift may be evaluated inserting Eqs.(\ref{eq: nertm}) and (\ref{eq: tertm}) into Eqs.(\ref{eq:sze5bis}) and (\ref{eq: defycomp}) thus obtaining\,:

\begin{equation}
\frac{\Delta T}{T_0} = - \frac{2 k_B \sigma_T T_{e_0} \,
n_{e_0}}{m_e c^2} \ {\times} \ \eta \ , \label{eq:sze7}
\end{equation}
with $\eta$ the so called {\it structure integral} defined as\,:

\begin{equation}
\eta = 2 \int_0^l {n_e(s)\over n_{e_0}}{T_e(s)\over T_{e_0}} dl'
\label{eq:sze8}
\end{equation}
which, for a given density and temperature profile, depends only on the geometry and the extension of the cluster along the line of sight. In Eq.(\ref{eq:sze8}), $l$ is the maximum extension of the gas along the line of sight. Measuring the lengths in units of the virial radius, it is $l = 1$ since the RTM model is truncated at $R_v$. Note that, usually, one takes $l \rightarrow \infty$ thus introducing a systematic bias whose effect we will examine later. A simple geometrical argument converts the integral in Eq.(\ref{eq:sze8}) in angular form introducing the angular diameter distance, $D_A$, to the cluster\,:

\begin{eqnarray}
\eta & = & 2 \theta_v D_A  \int_0^1{n_e(\chi) T_e(\chi) dl'} \nonumber \\
~ & = & 2 \theta_v D_A \int_{0}^{\sqrt{1 + \xi^2}}{n_e(\chi)
T_e(\chi) \frac{\chi d\chi}{\sqrt{\chi^2 - \xi^2}}} \label{eq:
etaint}
\end{eqnarray}
with $\chi= (l^2 + x^2 + y^2)/R_v^2$, and $\xi^2 = (x^2 + y^2)/R_v^2$. The integral in Eq.(\ref{eq: etaint}) can be further simplified introducing the auxiliary variable

\begin{displaymath}
\tilde{\alpha} = \frac{1 + s^2}{1 + \sqrt{\xi}} \ .
\end{displaymath}
The structure integral $\eta$, once evaluated, allows to calculate the comptonization parameter $y$, and then the temperature shift $\Delta T/T$, according to the formula (\ref{eq:sze5bis}). Adopting a flat $\Lambda$CDM model with $(h, \Omega_m) = (0.7, 0.3)$ and typical values for the virial radius and the virial temperature ($R_v \simeq 2 h^{-1} \ {\rm Mpc}$, $T_v \simeq 8 - 9 \ {\rm keV}$), we get a central number density $n_{e,0} = 5.07 \ {\rm cm^{-3}}$ and a shift in temperature at the cluster centre $\Delta T \simeq 9.3 \ {\rm mK}$.

\begin{figure}
\centering \resizebox{5.5cm}{!}{\includegraphics{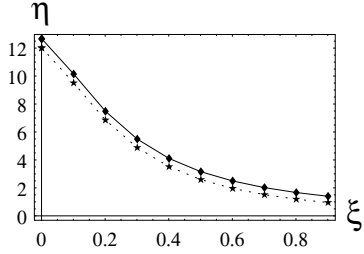}} \hfill
\caption{The structure integral $\eta$ vs $\xi$. The upper curve has been obtained by fixing $l = 1$ in Eq.(\ref{eq:sze8}), while the lower one refers to the case $l \rightarrow \infty$ .} 
\label{geo}
\end{figure}

\begin{figure}
\centering \resizebox{5.5cm}{!}{\includegraphics{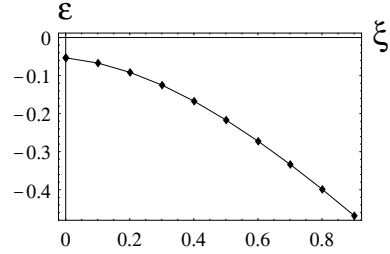}}
\hfill \caption{The relative error $\varepsilon$ on the structure integral $\eta$ vs $\xi$. It is\,: $\varepsilon = [\eta(l \rightarrow \infty) - \eta(l = 1)]/\eta(l = 1)$. } 
\label{error}
\end{figure}

Let us now investigate in more detail how the peculiarities of the model affect the structure integral (and hence the comptonization parameter and the temperature shift). First, we consider the effect of the finite extension of the RTM model. To this aim, in Fig.\,\ref{geo}, we plot the structure integral $\eta$ as function of $\xi$ with the upper (lower) curve obtained assuming $l = 1$ ($l \rightarrow \infty$) in Eq.(\ref{eq:sze8}). It is worth noting that the usual hypothesis of infinite extension may lead to significantly underestimate the SZE effect of the model by an amount that depends on the value of $\xi$. However, since what is usually measured is the temperature shift at the cluster centre, the relative error  is not dramatic being less than $\sim 10\%$ for $\xi < 0.2$ as it is shown in Fig.\,\ref{error}. Note, however, that the error due to the finite cluster extension for the RTM model is lower than the corresponding one for the standard $\beta$ model (\cite{jetzer}).

\begin{figure}
\centering \resizebox{5.5cm}{!}{\includegraphics{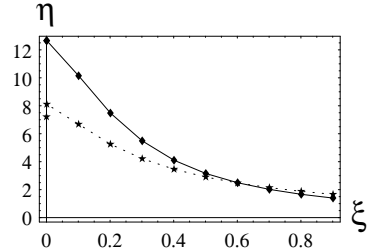}}
\hfill 
\caption{The structure integral $\eta$ vs $\xi$ for the RTM model using the correct temperature profile (upper curve) or the isothermality hypothesis (lower curve).} 
\label{iso}
\end{figure}

The temperature profile of the RTM model is approximately isothermal up to $\sim 0.2 R_v$. Moreover, the isothermality hypothesis is often used in SZE computations. It is thus interesting to investigate what is the systematic error induced by the simplifying assumption $T(r) = T_{e,0}$ for the RTM model. In this case, the structure integral may be analytically expressed in terms of hypergeometric functions\,:

\begin{equation}
\eta_{iso}(s) = 2 \theta_v D_A {\cal{Z}}(s) \label{eq: isoeta}
\end{equation}
with

\begin{eqnarray}
{\cal{Z}}(s) & = & \frac{1}{s^{1.5}} \left ( 1.19814 _2F_1 \left [ \{1.75, 1.75\}; \{0.5\}; \frac{0.09}{s^2} \right ] \right ) \nonumber \\
~ & ~ &  - \frac{1}{s^{2.5}} \left ( 0.65514 _2F_1\left [ \{2.25,
1.25\}; \{1.5\}; \frac{0.09}{s^2} \right ] \right ) \nonumber \ .
\label{eq: defh}
\end{eqnarray}
Fig.\,\ref{iso} shows that there is a dramatic change in the structure integral (and thus in the SZE temperature shift) for the RTM model if we use an isothermal temperature profile instead of that given by Eq.(\ref{eq: tertm}). Actually, even if the RTM model has an almost isothermal core up to $r = 0.2 \,R_v$, assuming an isothermal profile leads to a large error in the the structure integral. In particular, in the inner region of the cluster, where the SZE effect is measured, $\eta_{iso}$ is more than $30\%$ lower than the true $\eta$ thus leading to a similar error on the temperature shift. Moreover, this error is larger than the one due to the finite extension of the cluster that dominates in the outer regions.

\section{Comparison with {\it standard} gas profiles}

The peculiar non\,-\,isothermal temperature profile of the RTM model makes it different from most of the parametrizations used to describe the gas properties in galaxy clusters. It is thus particularly interesting to compare the SZ signal for the RTM model with that from more {\it standard} gas profiles. As interesting examples, we will consider the $\beta$ and the NFW models.

\subsection{The $\beta$\,-\,model}

The X\,-\,ray surface brightness in galaxy clusters is commonly fitted using the so called $\beta$\,-\,model (\cite{cff76}) whose density profile is\,:

\begin{equation}
\frac{\rho}{\rho_b} = \rho_{0 \beta} x_c^{-3\beta} \left ( 1 + \frac{x^2}{x_c^2} \right )^{-\frac{3 \beta}{2}}
\label{beta}
\end{equation}
where $x_c = r_c/R_v$ is the core radius in units of the virial radius, $\rho_{0 \beta}$ is related to the central electron number density $n_{e 0} = \rho_b \rho_{0 \beta}/(\mu m_p)$ and $\beta$ is a fitting parameter lying in the range $0.5 \le \beta \le 1$. The comptonization parameter for such a model is\,:

\begin{equation}
y = \frac{2 k_B \sigma_T}{m_e c^2} \int_{0}^{L}{n_e T_e dl} = \frac{k_B \sigma_T n_{e 0} T_{e 0}}{m_e c^2} \eta_{\beta}
\label{ybeta}
\end{equation}
with $\eta_{\beta}$ the structure integral given as\,:

\begin{eqnarray}
\eta_{\beta} & = & R_v x_c^{1 - \beta} \int_{\tilde{\xi}}^{\tilde{\xi} + \tilde{\xi}_L}
{\frac{(1 + \tilde{\xi}')^{-3 \beta/2}}{\sqrt{\tilde{\xi}' - \tilde{\xi}}} d\tilde{\xi}'} \nonumber \\
~ & = & R_v x_c^{1 - 3 \beta} \left ( 1 + \tilde{\xi}_L \right )^{-3 \beta/2} \ \times \nonumber \\
~ & \times & 
{_2F_1}\left [ \left \{ \frac{1}{2}, \frac{3}{2} \beta \right \}; \left \{ \frac{3}{2} \right \}; - \frac{1}{x_c^2(1 + \tilde{\xi}_L)} \right ] \ ,
\label{etabeta}
\end{eqnarray}
where we have defined $\tilde{\xi} \equiv x^2/x_c^2$. In order to compare the RTM and $\beta$ SZE signal we evaluate the quantity $\epsilon_y= 1 - y_{\beta}/y_{RTM}$. To this aim, it is convenient to first reparametrize the RTM structure integral as follows\,:

\begin{figure}
\centering
\resizebox{5.5cm}{!}{\includegraphics{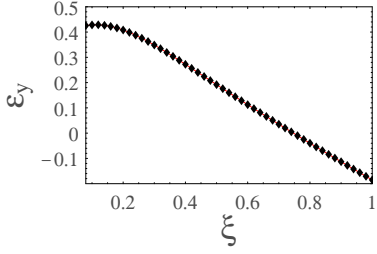}}
\hfill 
\caption{The relative error on the SZE signal induced by using the best fit $\beta$\,-\,model instead of the correct RTM one as function of the dimensionless radius $\xi = r/R_v$.}
\label{deltaeta_rtmbeta}
\end{figure}

\begin{eqnarray}
\eta_{RTM} = R_v x_{p_1}^{-1.5} \int_{\chi}^{\chi + \chi_L}
{\frac{\chi'^{0.008} \left ( 1 + \sqrt{\chi'} \right )^{-2.5} d\chi'}{\sqrt{\chi' - \chi} \left ( x_{p_2}^4 + \chi'^2 x_{p_1}^4 \right )^{0.13}}}
\label{etartm}
\end{eqnarray}
with $x_{p_1} = 0.04$, $x_{p_2} = 10^{-0.51}$ and $\chi = x^2/x_{p_1}^2$. With these settings, we get\,:

\begin{eqnarray}
\epsilon_y & = 1 - & \frac{\left ( 1 + \tilde{\xi} \right )^{- 3 \beta/2}}{x_{p_1}^{-1.5} x_c^{3 \beta - 1}} \  
{_2F_1}\left [ \left \{ \frac{1}{2}, \frac{3}{2} \beta \right \}; \left \{ \frac{3}{2} \right \}; - \frac{1}{x_c^2(1 + \tilde{\xi}_L)} \right ] \nonumber \\ 
~ & ~ & \times \ \left ( \int_{\chi}^{\chi + \chi_L}
{\frac{\chi'^{0.008} \left ( 1 + \sqrt{\chi'} \right )^{-2.5} d\chi'}{\sqrt{\chi' - \chi} \left ( x_{p_2}^4 + \chi'^2 x_{p_1}^4 \right )^{0.13}}} \right )^{-1} \ .
\end{eqnarray}
To estimate $\epsilon_y$, we have first to choose reasonable values for the $\beta$\,-\,model parameters. To this aim, we first set $\rho_{0 \beta} = \rho_{0}$, with $\rho_0$ the characteristic density of the RTM model, and then choose $(\beta, x_c)$ by fitting the $\beta$\,-\,model to the RTM density profile. We obtain as best fit values $(\beta, x_c) = (0.74, 0.04)$ in qualitative agreement with $(\beta, x_c) = (0.7, 10^{-1.53})$ given in Rasia et al. (2004). With this choice of parameters, the two profiles agree quite well, within few $\%$, in the inner regions, while the relative error increases up to $\sim 10\%$ in the outer zone. Nonetheless, due to the radically different temperature profiles, the SZE signal is quite different as can be seen in Fig.\,\ref{deltaeta_rtmbeta}. Actually,  $\epsilon_y$ turns out to be significantly different from unity even in the region $\xi \le 0.8$ where the two models fit each other quite accurately. In particular, we find that the SZE signal is larger for the RTM than for the $\beta$\,-\,model everywhere but in the extreme outer regions of the cluster where the situation is reversed. 

Actually, the most interesting implications of the non isothermal temperature profile for the RTM model with respect the standard $\beta$\,-\.model concerns the {\it detectability} of a cluster SZE signal, and therefore the statistics of the Sunyaev\,-\.Zeldovich clusters in different cosmological models. Even if all of the physics of the effect is coded in the Compton $y$ parameter, it is the total flux density from the cluster that is requested from the observational point of view. This is found by integrating the comptonization parameter over all the cluster face obtaining\,:

\begin{equation}
Y = \int{y(\vec{\theta}) d^2\vec{\theta}}
\label{detection}
\end{equation}
being $\vec{\theta}$ the angular position on the sky. Since $y$ is dimensionless, $Y$ is effectively a solid angle. A caveat is in order here. Any  SZE clusters survey (as for instance Plank) has some fixed angular resolution, which  will not allow to spatially resolve low mass clusters, and  even high mass cluster can be barely resolved (\cite{Aghanim97}). Therefore, a background $y_b$ Compton parameter will be present and will be dominated by low mass clusters since their higher number density overcompensates their lower individual contributions. It is also worth noting that an isotropic background would not matter since it could be completely removed. Because of this noise, the detectability of a SZ cluster will depend on the average background fluctuations that can be estimated as (\cite{bartelmann})\,:

\begin{equation}
\bar{y} = \int{y(\vec{\theta'}) b(\vec{\theta} - \vec{\theta'}) d^2\vec{\theta}'}
\label{detection2}
\end{equation}
being $b(\vec{\theta})$ the cluster beam profile. A cluster is assumed to be detectable if its integrated, beam\,-\,convolved Compton $y$ parameter is sufficiently large, i.e.\,:

\begin{equation}
\bar{Y} = \int{\bar{y}(\vec{\theta}) d^2\vec{\theta}} \ge \bar{Y}_{min}
\end{equation}
where the integral has to be evaluated over the area where the integrand sufficiently exceeds the background fluctuations. However, in the following, we will neglect for simplicity the background noise, i.e. we will simply consider $Y$ rather than $\bar{Y}$. Even within such a simplified situation, we will discuss some aspects of the SZE detectability of RTM clusters, which can be easily generalized in presence of the background noise. 

If the gas temperature profile is isothermal, the integrated SZE flux calculated according to Eq.(\ref{detection}) may be simply related to the cluster temperature weighted mass divided by $D_A^2$, being $D_A$ the angular diameter distance. Actually, in an isothermal regime, being $d\Omega = dA/D_A^2$, Eq.(\ref{detection}) becomes\,:

\begin{equation}
Y = \int{y(\vec{\theta}) d^2\vec{\theta}} \propto \frac{N_e \langle T_e \rangle}{D_A^2} \propto \frac{M \langle T_e \rangle}{D_A^2}
\label{detectioniso}
\end{equation}
being $N_e$ the total number of electrons in the cluster, $\langle T_e \rangle$ the mean electron temperature (which appears in an isothermal profile), and $M$ the total mass of the cluster (or the gas mass $M_{g}= f_{g} M$). From Eq.(\ref{detectioniso}), it turns out that an SZE survey detects all clusters above some mass threshold which thus has a crucial role for the estimate of the SZE cluster counts and its cosmological and astrophysical applications (\cite{carl2002}). 

\begin{figure} 
\centering
\resizebox{5.5cm}{!}{\includegraphics{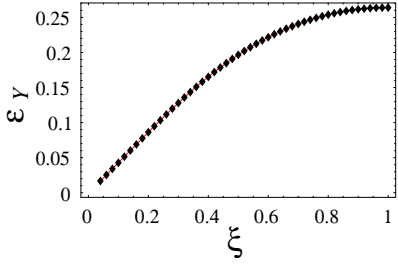}} 
\hfill
\caption{Dependence of $\epsilon_Y $ on the distance from the cluster center (in units of $R_v$).}
\label{delta_YRTMiso}
\end{figure}

However, for the RTM model, the temperature profile is approximately isothermal only in the inner regions ($r \le 0.2 R_v$) so that the question of the detectability has to be significantly revisited. As a first step, let us compare $Y_{RTM}$, the exact (i.e. computed using the correct temperature profile) integrated SZE flux for the RTM model, with $Y_{RTM}^{iso}$ that is evaluated under the isothermal approximation. These quantities are given as\,:

\begin{equation}
Y_{RTM} \propto 4 \pi \int_{0}^{X}{\frac{x^{2.016}}{(x_{p_1} + x)^{2.5} (x_{p_2}^4 + x^4)^{0.13}} dx} \ ,
\label{detrtm}
\end{equation}

\begin{equation}
Y_{RTM}^{iso} \propto 5.33 ({\cal{Y}} - x_{p_1}^{1/2})
\label{detrtmiso}
\end{equation}
with\,:
\begin{equation}
{\cal{Y}} =
\frac{x_{p_1}^3 + X \left [ x_{p_1} \left ( 2.5 x_{p_1} + 1.875 X \right ) + 0.375 X^2 \right ]}
{(x_{p_1} + x)^{2.5}} \ .
\label{eq: detrtmiso}
\end{equation}
with $X = r/R_v$. Fig.\,\ref{delta_YRTMiso} shows how $\epsilon_Y = 1 - Y_{RTM}^{iso}/Y_{RTM}$ increases with the distance from the cluster centre being of order $25\%$ at the virial radius. As a result, Eq.(\ref{detectioniso}) for a RTM cluster is replaced by\,:

\begin{equation}
Y \propto \frac{M_{eff} \langle T_e \rangle}{D_A^2} = \frac{\nu M \langle T_e \rangle}{D_A^2}
\label{detectionrtmexact}
\end{equation}
being $\nu = Y_{RTM}/Y_{RTM}^{iso}$. Because of the coefficient $\nu > 1$, the SZE detectability of a RTM cluster is naturally improved\,: for a fixed threshold value $Y_{lim}$, it is possible to detect less massive clusters than in the isothermal case. 

\begin{figure} 
\centering
\resizebox{5.5cm}{!}{\includegraphics{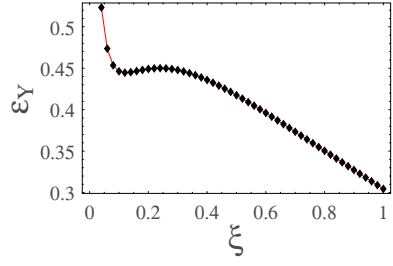}} 
\hfill
\caption{Comparison of the integrated SZE flux of a $\beta$ and RTM model with $\epsilon_{Y} = 1 - Y_{\beta}/Y_{RTM}$ plotted as function of the dimensionless distance from the cluster centre.}
\label{det_rtm_beta}
\end{figure}

As a consequence, a RTM cluster is also detectable at lower mass regimes than a $\beta$\,-\,model cluster, as shown from a direct comparison. To this aim, we first remind tha the integrated SZE flux for a $\beta$\,-\,model is\,:

\begin{equation}
Y_{\beta} \propto \frac{\xi^2}{3} \times {_2F_1}\left [ \left \{ \frac{3}{2}, \frac{3}{2} \beta \right \}; 
\left \{ \frac{5}{2} \right \}, - \frac{\xi^2}{x_c^2} \right ] \ .
\end{equation} 
In Fig.\,\ref{det_rtm_beta}, we plot the relative deviation of the best fit $\beta$\,-\,model and the corresponding RTM model. It turns out, indeed, that the integrated SZE flux is rather higher for a RTM cluster.

\subsection{The NFW model}

Even if it is mostly used to describe the dark matter rather than the gas distribution in galaxy clusters, it is nonetheless interesting to compare the SZE predictions for the RTM model with the same quantities evaluated for the NFW model. 

\begin{figure} 
\centering
\resizebox{5.5cm}{!}{\includegraphics{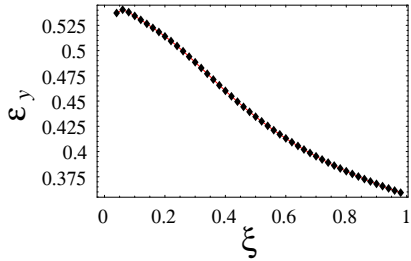}} 
\hfill
\caption{Relative discrepancy between the Componization parameter for a RTM model and its best fitting NFW model.}
\label{rely_rtmnfw}
\end{figure}

\begin{figure} 
\centering
\resizebox{5.5cm}{!}{\includegraphics{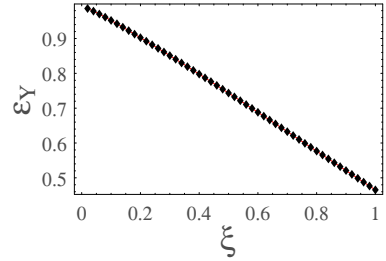}} 
\hfill
\caption{Relative discrepancy between the integrated SZE fluxes for a RTM model and its best fitting NFW model.}
\label{deltaY_rtmnfw}
\end{figure}

As a first step, we fit the NFW model to the RTM one obtaining $c_{NFW} = 5.99$ as best fit value for the concentration parameter. In agreement with Rasia et al. (2004), we find that the NFW model slightly overestimates the gas density in the range $0.04 \le x \le 0.4$, while it works sufficiently well in the outer regions. Assuming an isothermal profile for the gas temperature, the comptonization parameter for the NFW model turns out to be\,:

\begin{eqnarray}
y_{NFW} & \propto & \left [ \left ( X^2 - x_s^2 \right ) \left ( x_s + \sqrt{L^2 + X^2} \right ) \sqrt{x_s^4 - x_s^2 X^2} \right ]^{-1} \nonumber \\
~ & \times & x_s^3 \left \{ L \sqrt{x_s^2 - X^2} + x_s^2 \left ( 1 + \frac{\sqrt{L^2 + X^2}}{x_s} \right ) \ \times \right . \nonumber \\
~ & ~ & \left . \ \ \ \ \ \ln{\left [ \frac{\left ( X^2 - x_s^2 \right ) {\cal{Y}}_1}{{\cal{Y}}_2 {\cal{Y}}_3} \right ]} \right \} \ ,
\label{ynfw}
\end{eqnarray}
with\,:

\begin{equation}
{\cal{Y}}_1 = x_s \left ( X^2 + x_s \sqrt{L^2 + X^2} \right ) - L \sqrt{x_s^4 - x_s^2 X^2} \ ,
\end{equation}

\begin{equation}
{\cal{Y}}_2 = \frac{X \left (X^2 - x_s^2 \right )}{x_s \sqrt{x_s^4 - x_s^2 X^2}} \ ,
\end{equation}

\begin{equation}
{\cal{Y}}_3 = x_s^3 \left ( x_s + \sqrt{L^2 + X^2} \right ) \sqrt{x_s^2 - X^2} \ ,
\end{equation}
having denoted with $x_s = 1/c_{NFW}$, $L$ the cluster length along the line of sight and $X = r/R_v$. In Fig.\,\ref{rely_rtmnfw}, we compare the SZE signal predicted from both models, plotting the relative discrepancy $\epsilon_y = 1 - y_{NFW}/y_{RTM}$ between the comptonization parameters. We see that the SZE signal from a RTM cluster exceeds that from a NFW model as yet observed when comparing to the $\beta$\,-\,model. The RTM model thus emerges as the most effectively detectable also at larger distances from the cluster centre. This is yet more clear from Fig.\,\ref{deltaY_rtmnfw} where we compare the integrated SZE flux for the RTM and NFW models.

\section{A short comment on possible systematic errors}

There is a potential caveat about the RTM model that could affect the main results we have discussed insofar. The simulations that have been considered by Rasia et al. (2004) to develop the model are based non radiative hydrodynamics so that cooling flows and cold blobs that may eventually be along the line of sight are not reproduced. Taking into account these effects is a difficult task, but we expect, qualitatively, that cooling flows and conduction should lower the central temperature and thus increase the SZE signal at the centre. However, we stress that detailed simulations also taking into account star formation and feedback processes are needed to investigate how the gas temperature profile (and thus the SZE signal) are affected. Some preliminary results have been presented (\cite{DJSBR04}), but the effect on the SZE flux has still to be investigated.

As a final remark, it is worth noting that merging of clusters has not been considered, but it is likely that this does not affect the main results. Actually, the sample of simulated clusters analyzed by Rasia et al. (2004) comprises both relaxed, unrelaxed and post\,-\,merging systems and the RTM model turns out to be a good fit to the full sample which is an evidence strongly suggesting that merging effects do not alter significantly the cluster structure. As a result, the lensing properties of the model are likely to be affected only when the merging is in progress in which case an external shear could mimic to first order the deviations from spherical symmetry of the outer regions of the dark matter halo. A stronger effect is expected for the impact of merging on the SZE signal and the X\,-\,ray emission, since they depend on $n_{e_0}$ and $n_{e_0}^2$ respectively. Although further detailed simulations are needed to quantitatively address this question, some partial analytical results have been already obtained for some special mergers regimes, when the presence of cold fronts marks the late merging stages\,: namely the transonic and the subsonic mergers. It turns out that in the transonic regime the frequency spectrum of the SZE signal $g(p)$ changes, due to a shock particle re\,-\,acceleration mechanism, depending on the concentration, which induces a new electron population. As net effect, the crossover changes still up to $\sim 10\%$. In the subsonic case, instead, $g(p)$ remains unchanged, but the amplitude of the SZE signal is enhanced in a not negligible way, mainly in the interior regions of the cluster where it reaches also $\sim 30 - 40 \%$  (\cite{koch04}). Moreover, it is well known (\cite{torri04}) that the X\,-\,ray luminosity overall increases during merging so that it is likely that, as net effect, the SZE signal is enhanced during cluster merging events.

\section{Conclusions}

Being detectable at high redshift, galaxy clusters are promising tools for determining cosmological parameters and testing theories of structures formation. Hydrodynamical simulations are able to predict not only the dark matter mass distribution, but also the density law and the temperature profile of the gas component thus allowing a study of both the lensing properties and the Sunyaev\,-\,Zel'dovich effect due to the cluster. This has been the aim of the present paper where we have applied this study to the RTM model, recently proposed by Rasia et al. (2004) on the basis of the results of a large set of high resolution hydrodynamical simulations.

Assuming spherical symmetry in the density profile, we have derived the main lensing properties of the RTM model evaluating the deflection angle and the lensing potential that allows us to write down the lens equations. Most of the work has been devoted to a detailed investigation of the critical curves structure of the RTM model since it is the position of radial and tangential arcs (that forms just near the critical curves) that gives the most useful constraints on the cluster parameters. The main results are summarized below. 

\begin{enumerate}

\item{The RTM model always gives rise to both radial and tangential arcs, but their distance from the cluster centre is comparable to that observed in real systems only for values of the concentration much higher than typically predicted by the numerical simulations of Rasia et al. (2004).}

\item{Adding a giant elliptical galaxy (described with the Hernquist profile) to the cluster potential increases the distance between the radial and the tangential critical curves and allows to form observable radial arcs for lower values of the cluster concentration.}

\item{The shape of the critical curves may be significantly altered by a perturbing shear (mimicking an internal ellipticity or the effect of tidal perturbations) thus changing the positions of both radial and tangential arcs by an amount that can be up to $\sim 30\%$ for a shear strength $\gamma = 0.15$ (see Table 1).}

\item{Fitting an NFW model to a cluster that is intrinsically described by the RTM model leads to overestimate the cluster mass by an amount that depends on the concentrations $c_{RTM}$ and $c_{NFW}$ used in the modeling.}

\end{enumerate}
The work presented should be complemented by a detailed study also taking into account the presence of substructures (predicted by the CDM paradigm of structures formation) in the dark halo cluster. The total lensing potential thus should be made of the sum of the contributions from the elliptical RTM model, the external shear (due, e.g., to tidal perturbations or large scale structure), the bright cluster galaxy and the satellite haloes. This approach is quite complicated given the high number of unknown parameters entering the modeling, but one could resort to X\,-\,ray data to constrain (at least) the RTM parameters $(M_v, x_p)$.  However, the most compelling test is a direct comparison with real systems showing tangential and radial arcs. To this aim, adding the BCG contribution (and eventually the shear term) to the RTM lensing potential should give a sufficiently accurate cluster model to be compared with the data on the arcs position. Having determined the galaxy scalelength and surface brightness from photometric observations, this method should allow us to test whether the RTM model may reproduce the observed arcs positions and to constrain its parameters. This will be performed in a future paper.  

One of the most interesting feature of the RTM model is that the temperature profile is approximately isothermal only up to $0.2 \ R_v$. We have investigated the implications of the RTM model for the intracluster gas on the SZE evaluating the structure integral that determines the temperature decrement. We have thus taken into account both the finite extension of the model and its peculiar temperature profile estimating the errors induced by the usually adopted simplifications of infinite extension and isothermal temperature. The main results are as follows.

\begin{enumerate}

\item{Neglecting the finite extension of the cluster systematically underestimates $\Delta T/T$ by an amount that is less than $10\%$ in the inner regions of the cluster so that it may be neglected in a first order analysis.}

\item{Using $T = T_ {e,0}$ as temperature profile instead of that found by RTM leads to underestimate $\Delta T/T$ up to $30\%$ (in absolute value) in the centre.} 

\item{The comptonization parameter $y$ for the RTM model is higher than that of both the $\beta$ and the NFW models even if the parameters are chosen in such a way that the gas density is well fitted by the three models. Using the best fitting $\beta$ (NFW) model instead of the correct RTM one underestimates $y$ up to $\sim 40\%$ ($\sim 52\%$, respectively) in the inner cluster regions.}

\item{The non isothermal temperature profile leads to an integrated SZ flux which is higher for the RTM model than for both the $\beta$ and NFW models by an amount that depends on the distance from the cluster centre, but can be as high as $\sim 45\%$ and $\sim 90\%$ for the $\beta$ and NFW models respectively. As a result, less massive clusters should be detected in SZE survey if the RTM model is indeed the correct one.}

\end{enumerate}
As a final remark, we would like to stress that, in our opinion, a combined analysis (from the theoretical and observational point of view) of the both the lensing properties and the SZ temperature decrement could be the best method to validate a given cluster model independently on the peculiarities of the numerical simulations inspiring it.

\begin{acknowledgements}
We thank Elena Rasia for the interesting discussions on the RTM model characteristics and the impact of merging and cooling flows on the SZ signal. P. Koch is acknowledged for the interesting discussion about the transonic and subsonic merger regimes. We are also grateful to G. Longo, C. Rubano and M. Sereno for a careful reading of the manuscript. Finally, the authors are indebted with the referee, Percy Gomez, for his constructive report that has helped to improve the paper.
\end{acknowledgements}

\end{document}